\documentclass[11pt,a4paper]{article}

\usepackage{amsfonts}
\usepackage{amssymb}
\usepackage{graphicx}
\usepackage{amsmath}
\usepackage[dvipdfm]{hyperref}
\usepackage{enumerate}

\RequirePackage{mathrsfs} \RequirePackage[sc]{mathpazo}
\RequirePackage{wasysym} \RequirePackage{setspace}
\begin{document}

\title{On Thermodynamical  Behaviors of  Kerr-Newman AdS Black Holes}
\author{A. Belhaj$^{1,2}$, M. Chabab$^{2}$, H. El Moumni$^{2}$, L. Medari$^2$, M. B. Sedra$
^{3}$ \\
\\
{\small $^{1}$ D\'epartement de Physique, Facult\'e
Polydisciplinaire, Universit\'e Sultan
Moulay Slimane, B\'eni Mellal, Morocco} \\
{\small $^{2}$ High Energy Physics and Astrophysics Laboratory,
FSSM, Cadi
Ayyad University, Marrakesh, Morocco } \\
{\small $^{3}$ Universit\'{e} Ibn Tofail, Facult\'{e} des Sciences, D\'{e}%
partement de Physique, LHESIR, K\'{e}nitra, Morocco.} }
\date{\today}
\maketitle

\begin{abstract}
We reconsider the study of  critical behaviors of
Kerr-Newman AdS black holes in four dimensions.  The study  is made
in terms of the moduli space parameterized by the charge $Q$  and
the rotation parameter $a$,  relating  the mass $M$  of the black
hole and  its
 angular momentum $J$  via the  relation
$a=J/M$. Specifically, we discuss  such  thermodynamical
behaviors in the presence of a positive cosmological constant
considered  as a thermodynamic pressure and its conjugate quantity
as a thermodynamic volume.  The equation of state for a charged
RN-AdS black hole predicts a critical universal number depending on
the $(Q,a)$ moduli space.   In the vanishing limit of the $a$
parameter, this prediction recovers the usual  universal  number in
four dimensions.  Then, we find the bounded regions of the moduli
space allowing the consistency of the model with real
thermodynamical variables.

\end{abstract}

\newpage

Four-dimensional black holes have received  increasing attention
in the context of supergravity theories embedded  either in
superstrings or in  M-theory compactified on internal
spaces\cite{VS,S}. This involves  Reissner-Nordstrom (RN) black holes
being static, spherically symmetric configurations minimizing the
Maxwell--Einstein action.   These  solutions  are completely defined
by giving two parameters: the charge of the black hole $Q$ and the
mass $M$.  Such parameters have been  explored to  fix some stringy
moduli using the attractor mechanism \cite{FKS1,FKS2}.
\\
\par  Recently, intensive efforts have been devoted to investigation of
the thermodynamic behaviors of the black hole backgrounds including
Reissner--Nordstrom Anti-de-Sitter (RN-AdS) black solutions in four and  in
various dimensions\cite{our,KM,30,4,5,50,6,7,8}.  More precisely,
the  equation of state for  certain  black holes   has been worked
out  and  it has been realized that this analysis shares similar
feature as the Van der Waals $P-V$ diagram \cite{8,KM}.
\\
\par
The $P-V$ criticality of RN-AdS black holes with spherical
configurations have been extensively investigated using the same
method. In this way, the behavior of the Gibbs free energy in the
fixed charge ensemble has been dealt with. In particular, the phase
transition in the ($P,T$)-plane  has been studied  in \cite{KM}. In
fact,   a nice connection between the behavior of
the RN-AdS black hole system and the Van der Waals fluid has been shown. Moreover,
it has been realized that  $P$-$V$ criticality, Gibbs free energy,
first order phase transition and the behavior near the critical
points can be associated with  the liquid-gas system.
\\
\par
 Following the strategy of  \cite{KM}  to describe the   four dimensional case, we  have  discussed
   the critical behavior of the charged RN-AdS black holes in arbitrary dimensions of the
spacetime\cite{our}. Considering  the cosmological constant
$\Lambda$ as a thermodynamic pressure and its conjugate quantity as
a thermodynamic volume, we have given   a comparative study in terms
of the dimension and the displacement of the  critical points. These
parameters can be used to control the transition between the small
and the large black holes. More precisely,  it has  been shown  that
such behaviors vary nicely in terms of the dimension of the
spacetime in which the black holes reside. In an arbitrary dimension
$d$, we  have obtained   an universal number given by
\begin{equation}
\chi=\frac{P_c v_c}{T_c}=\frac{2d-5}{4d-8}
\end{equation}
connecting the space-time dimension $d$  with the  universal number
$\chi$. This equation recovers the four dimensional value
\begin{equation}
\chi=\frac{P_c v_c}{T_c}=\frac{3}{8}.
\end{equation}
\par
More recently,  the critical behaviors of  the charged and the
rotating AdS black holes   in the presence of the  electrodynamic
effects have been investigated in   \cite{GKM}. In four dimensions,
it has been found that neutral slowly rotating black holes involve
the same critical  behavior. Among others,  
the relation  $\chi=\frac{P_c v_c}{T_c}=\frac{5}{12}$ has been obtained. However, this
number,  in  fact,  should  depend on  the extra rotating parameter
and should recover the usual value (2) in the vanishing limit
of such a parameter.\\
\par The aim of this work is to contribute to these topics by
reconsidering the study of  the critical behaviors of the
Kerr-Newman AdS black holes. The  present study  is made in terms of
the moduli space parameterized by the charge $Q$  and   the rotation
parameter $a$. The latter  is given in terms of   the mass $M$  of
the black hole and its  angular momentum $J$  via   the expression
$a=J/M$.  In this way, these two parameters together with
the displacement of the  critical points can be used to control the
transition between the small and the large black holes. Among
others, we find  that the equation(2) has been modified. It will be
given in terms of the $Q$ and $a$ parameters. Then, we present
numerically the bounded regions of the moduli space allowing the
consistency of
the model with real thermodynamical  quantities.\\
\par  To go ahead, we consider the Einstein-Maxwell-AdS
action in four dimensions given by
\begin{equation}
\mathcal{I}=-\frac{1}{16\pi G}\int_{M}dx^{4}\sqrt{-g}\left[ R-F^{2}+2\Lambda %
\right]   \label{action}
\end{equation}
where  $F=dA$ is the field strength   where $A$ is the potential
1-form. Here, $\Lambda $  can be identified with  $-\frac{3}{\ell
^{2}}$ defining the cosmological constant associated with the characteristic length scale $%
\ell $.  The variation of  the above action with respect to the
metric tensor  provides  the Kerr-Newman AdS solution, given in
Boyer-Lindquist coordinate  by \cite{xm},
\begin{eqnarray}\label{ds}\nonumber
ds^{2}&=&-\frac{1}{\Sigma}[\Delta_r-\Delta_\theta a^2 sin^2\theta]dt^2-\frac{\Sigma}{\Delta_r}dr^2
+\frac{\Sigma}{\Delta_\theta}d\theta^2\\&+&\frac{1}{\Sigma\Xi^2}[\Delta_\theta(r^2+a^2)^2-\Delta_r a^2 sin^2
\theta]sin^2\theta d\phi^2
-\frac{2a}{\Sigma\Xi}[\Delta_\theta(r^2+a^2)-\Delta_r]sin^2\theta dtd\phi
\end{eqnarray}
where
\begin{equation}
\Sigma=r^2+a^2cos^2\theta,\quad\quad\Xi=1+\frac{1}{3}\Lambda a^2
\end{equation}
\begin{equation}
\Delta_\theta=1+\frac{1}{3}\Lambda a^2cos^2\theta,\quad\quad
\Delta_r=(r^2+a^2)(1-\frac{1}{3}\Lambda r^2) -2Mr+Q^2.
\end{equation}
In the negative values of  $\Lambda$, the horizons of the metric
$(\ref{ds})$  can be   derived from the  vanishing condition of the
following quantity
\begin{eqnarray}\nonumber
\Delta_r&=&(r^2+a^2)(1-\frac{1}{3}\Lambda r^2)-2Mr+Q^2\\
&=&-\frac{1}{3}\Lambda \left[r^4-\left(\frac{3}{\Lambda}-a^2\right)r^2
+\frac{6M}{\Lambda}r-\frac{3}{\Lambda}(a^2+Q^2)
 \right]\\
&=&-\frac{1}{3}\Lambda(r-r_{++})(r-r_{--})(r-r_+)(r-r_-).
\end{eqnarray}
In fact  the equation  $\Delta_r = 0$ has four  different roots,
 as shown in \cite{31}. $r_{++}$ and $r_{--}$ are a pair of complex conjugate
roots, while $r_+$ and $r_-$ are two real positive roots. Assuming
that $r_+
> r_-$,  the condition $r = r_+$  describes  the event horizon.\\
\par
Using the thermodynamical calculation  techniques \cite{32},  the
black hole temperature  reads as,
\begin{equation}
T=\frac{1}{\beta }=\frac{3r_+^4
+(a^2+\ell^2)r_+^2-\ell^2(a^2+Q^2)}{4\pi\ell^2 r_+(r_+^2+a^2)}.
\label{temperature}
\end{equation}
Following the analysis of  \cite{KM},  we can get the equation of
state for a Kerr-Newman AdS black hole $P=P(V,T)$ in four dimension.
For a generic point in the $(Q, a)$  moduli space, the calculation
leads to, 
\begin{equation}  \label{state}
P=\frac{3 \left(r_+ \left(4 \pi a^2
T+r_+ \left(4 \pi  r_+
   T-1\right)\right)+a^2+Q^2\right)}{8 \pi  r_+^2 \left(a^2+3
   r_+^2\right)}.
\end{equation}
In this expression, the event horizon radius $r_+$   is given by
\begin{equation}  \label{hor}
r_+=\left(\frac{3 V}{4}\right)^\frac{1}{3}
\end{equation}
 where $V$ is the thermodynamic volume of the black hole, being
 the  volume of  the four-dimensional sphere. It is worth noting that (\ref{state})  can be identified with the
  equation given in  \cite{GKM} by considering a particular normalization for   the mass and charge parameters.

\par  As made in  \cite{KM}, the
physical pressure and temperature  take the following forms
\begin{equation}
Press=\frac{\hbar c}{l_P}P, \;\;\; Temp=\frac{\hbar c}{k} T
\end{equation}
where the Planck length is given by $l_P^2=\frac{\hbar G_4}{c^3}$.
Multiplying (\ref{state}) by $\frac{\hbar c}{l_P}$,  one can get  a
compact expression for  the  $Press$  thermodynamical variable.
Indeed, the calculations give  the following form
\begin{equation}
Press=\frac{3 k r_+ \text{Temp}}{2 \left(a^2+3 r_+^2\right) \ell
   _p^2}+\ldots
\end{equation}
A close inspection around the Van der Waals equation given by $\left(P+\frac{%
a}{v^2}\right)(v-b)=k T$   has shown  that  we  can identify the
specific volume $v $ with
\begin{equation}
v=2 \ell_p^2 r_+.
\end{equation}
 In this way, the equation of state (\ref{state}) takes the
 following form
\begin{equation}
P=\frac{3 \left(a^2 (8 \pi  T \mathit{v}+4)+4 Q^2+\mathit{v}^2 (2
   \pi  T \mathit{v}-1)\right)}{8 \pi  a^2 \mathit{v}^2+6 \pi
   \mathit{v}^4}.\end{equation}
Having derived such quantities, we present a numerical  discussion on
the obtained results. These calculations  allow one to discuss the
corresponding $P$-$V$ diagram. It is plotted in the figure 1.

\begin{center}
\begin{figure}[tbp]
\begin{tabbing}
\hspace{8cm}\=\kill
\includegraphics[scale=1]{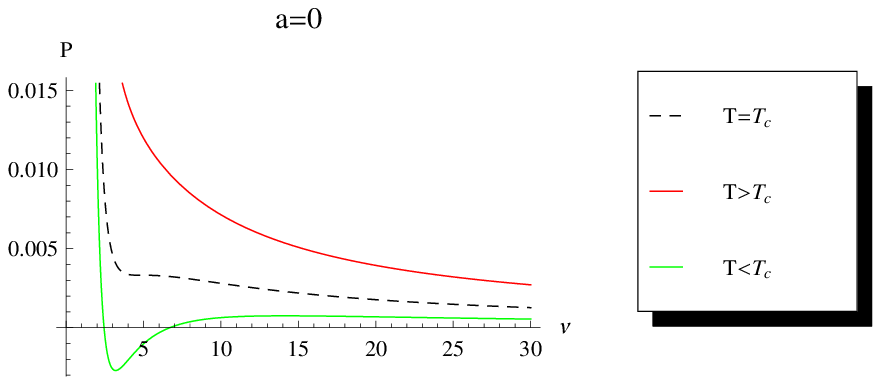}\\
\includegraphics[scale=1]{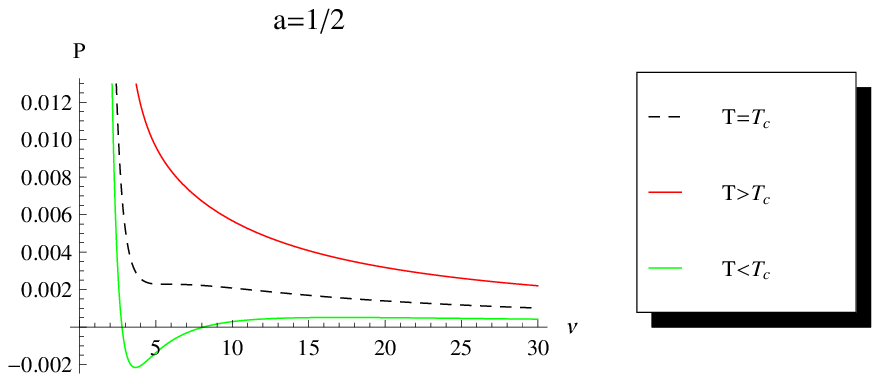} \\
\includegraphics[scale=1]{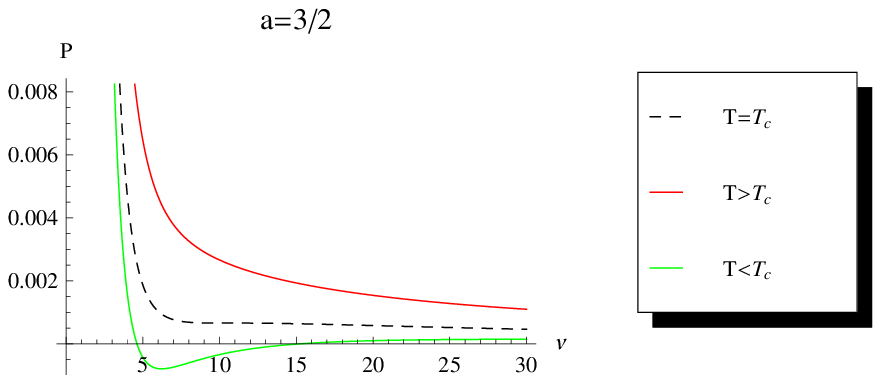}  
\end{tabbing}
\vspace*{-.2cm} \caption{The $P-V$ diagram of Kerr-Newman AdS black
holes, where $T_c$ is the critical temperature and the charge is
equal to $1$ } \label{fig1:fig2}
\end{figure}
\end{center}

It is seen from Fig. 1  that for $Q \neq 0$ and for $T< T_c$, the
behavior looks like an extended Van der Waals gas and the
corresponding system involves inflection points defining critical
points. These  points   satisfy the  conditions,
\begin{equation}
\frac{\partial P}{\partial v}=0,\;\;\;\; \frac{\partial^2 P}{\partial v^2}=0.
\end{equation}
After calculations, we obtain  the following  expressions
\begin{equation}
v_c=2 \sqrt{3 a^2+2 Q^2+\frac{32 a^4+41 a^2 Q^2+12 Q^4}{\sqrt[3]{3} X}+\frac{X}{3^{2/3}}}
\end{equation}
and
\begin{eqnarray}
T_c&=&\frac{3 \sqrt{Y} \left(8 Y \left(311 a^2+207 Q^2\right)+5424 a^2 \left(a^2+Q^2\right)-69 Y^2\right)}{3328
   \pi  a^6}
\end{eqnarray}
while the critical pressure  takes the form
\begin{eqnarray}
P_c=\frac{X \left(-2496 a^6 Z+3 \sqrt{3} a^2 X \text{Z'} \sqrt{\frac{Z}{X}}+7488 a^6 X
   \left(a^2+Q^2\right)+\sqrt{3} Z \text{Z'} \sqrt{\frac{Z}{X}}\right)}{6656 \pi  a^6 Z \left(a^2
   X+Z\right)}.
\end{eqnarray}
The quantities  $X$, $Y$, $Z$ and  $Z'$, appearing in the above
equations,  are given respectively by
\begin{equation}
\small
X=\sqrt[3]{312 a^6+600 a^4 Q^2+369 a^2 Q^4+\sqrt{3} \sqrt{-320 a^{12}-1152 a^{10} Q^2-1488 a^8 Q^4-809 a^6
   Q^6-153 a^4 Q^8}+72 Q^6}
\end{equation}
\begin{equation}
Y=\frac{128 a^4}{\sqrt[3]{3} X}+\frac{164 a^2 Q^2}{\sqrt[3]{3} X}+12 a^2+\frac{16\ 3^{2/3} Q^4}{X}+8
   Q^2+\frac{4 X}{3^{2/3}}
\end{equation}
\begin{equation}
Z=32\ 3^{2/3} a^4+a^2 \left(41\ 3^{2/3} Q^2+9 X\right)+12\ 3^{2/3} Q^4+6 Q^2 X+\sqrt[3]{3} X^2
\end{equation}

and

\begin{equation}
Z'=16272 a^4 \sqrt{Y}+16272 a^2 Q^2 \sqrt{Y}+7464 a^2 Y^{3/2}+4968 Q^2 Y^{3/2}-207 Y^{5/2}
\end{equation}
It is worth noting that  the vanishing condition of the  $a$
parameter  recovers  the result of \cite{KM}
\begin{equation}
T_{c_0}=\frac{1}{3 \sqrt{6} \pi  Q},\quad v_{c_0}=2 \sqrt{6} Q,
\quad  P_{c_0} =\frac{1}{96 \pi  Q^2},  \quad V_{c_0}=8\sqrt{6}\pi
Q^3.
\end{equation}

In what follows, we discuss  the effect  of   the parameter $a$
on  the locality of the critical points.  We plot in figure $2$ the
case of some particular points living in  the vertical line $Q=1$ of
the $(Q, a)$ moduli space. General study could be carried by varying
also the charge parameter.
\begin{center}
\begin{figure}[!h]
\center
\includegraphics[scale=1]{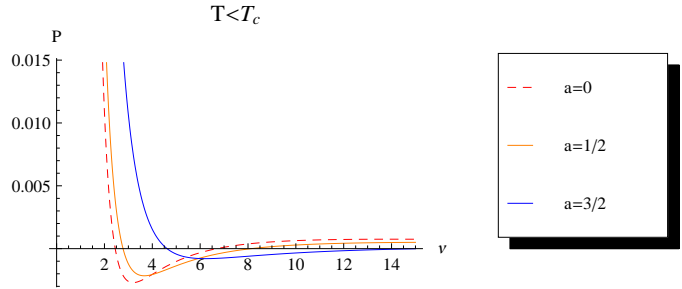}
\caption{The $P-V$ diagram of Kerr-Newman AdS black holes , for different values of the parameter $a$ and the charge is equal
to $1$ }
\label{fig1:fig2}
\end{figure}
\end{center}

It isfound form  figure 2   that  the parameter $a$ controls
 the position of the critical point change.  For  large value of  $a$,   the van der Waals
  behavior  tends to the ideal gas case.

After an examination  of  the  generic  regions   of  the $(Q, a)$
moduli space, we observe that the critical temperature is not
defined for all values of  $a$ and  $Q$.  At  the vicinity of the
origin  of the $(Q,a)$ moduli space, the temperature becomes  maximal.

 In the  small values of $a$,  the  critical coordinates
 can be reduced to,
\begin{eqnarray}
 T_c&=& \frac{1}{3 \sqrt{6} \pi  Q}-\frac{23 a^2}{72 \left(\sqrt{6} \pi  Q^3\right)}+\mathcal{O}\left(a^4\right)\\
  v_c&=& 2 \sqrt{6} Q+\frac{59 a^2}{6 \sqrt{6} Q}+\mathcal{O}\left(a^4\right)\\
   P_c&=& \frac{1}{96 \pi  Q^2}-\frac{11 a^2}{576 \left(\pi  Q^4\right)}+\mathcal{O}\left(a^4\right)
 \end{eqnarray}
 This  limit produces  a critical universal number
 \begin{equation}\label{universal}
 \chi=\frac{3}{8}-\frac{a^2}{48 Q^2}+\mathcal{O}\left(a^4\right)
 \end{equation}
This   expression has the following nice features:
\begin{itemize}
  \item  It  depends on the $(Q,a)$  moduli space.
 \item   It is  valid  only for charged  black holes.
 \item  For vanishing value of the $a$ parameter,  it  can be reduced to
 the usual value $\frac{3}{8}$  given in equation (2).\end{itemize}
 
  Moreover, we note  that  the  line  $(0, a)$ of the moduli
space $(Q, a)$ is singular corresponding to  an indefinite critical
point. In this singular line,  the equation of state  $(\ref{state})$ reduces to 
$P|_{Q=0}= \frac{3 \left(4 \pi  a^2 r_+ T+a^2+4 \pi
r_+^3 T-r_+^2\right)}{8 \pi  r_+^2 \left(a^2+3 r_+^2\right)}$
representing an ideal gas  behavior with absence of  the
critical  points as shown in the figure $4$. This singularity can be
removed by a taking a charged black hole.

At this level,  we   note that $(\ref{universal})$ recovers exactly
the result of  \cite{KM} by setting $a=0$. It produces also the
result   of \cite{our}  by considering four dimensional   black
hole. It should be interesting to see   the connection with  the
result of \cite{GKM}.  We  believe that this link deserves more study.  We decide to explore this  issue for future work.\\

\par
Extra  information on  the $(Q, a)$ moduli space can  be  derived by
studying the  critical thermodynamical variables. In what follows,
we will be interested in such quantities. The latter will shed
light on the physical   regions of the $(Q, a)$ moduli space in which
the thermodynamical variables have real values. In particular, we
give a numerical study. Indeed, the numerical result of  the
critical temperature and the  pressure are plotted in figure 3.
\begin{center}
\begin{figure}[!h]
\center
\begin{tabbing}
\hspace{9cm}\=\kill
\includegraphics[scale=.8]{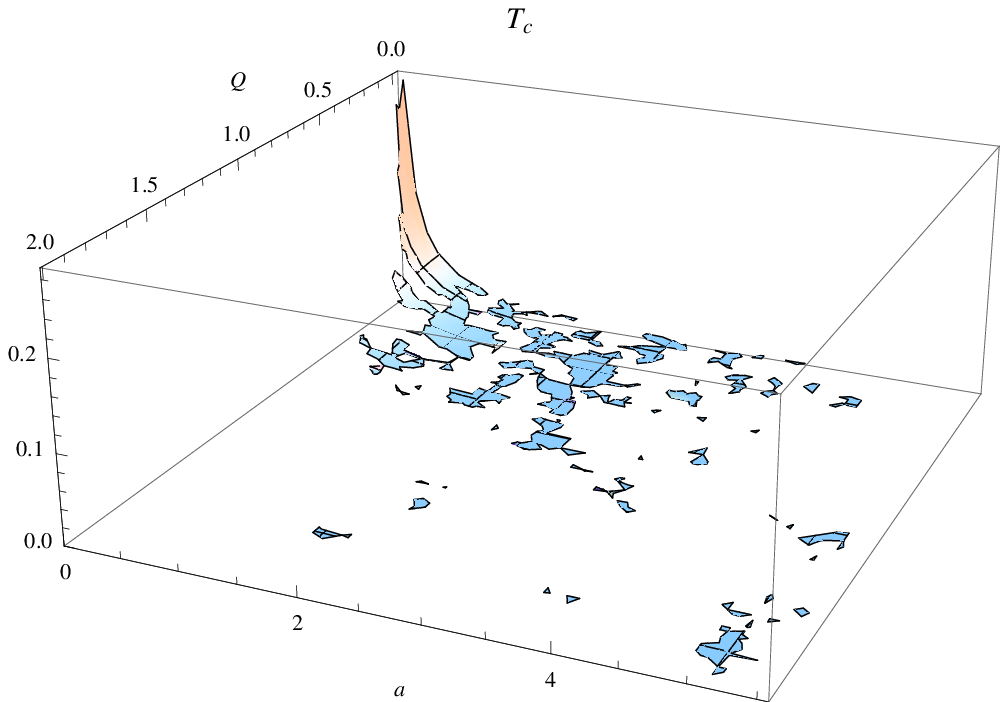} \> \includegraphics[scale=0.8]{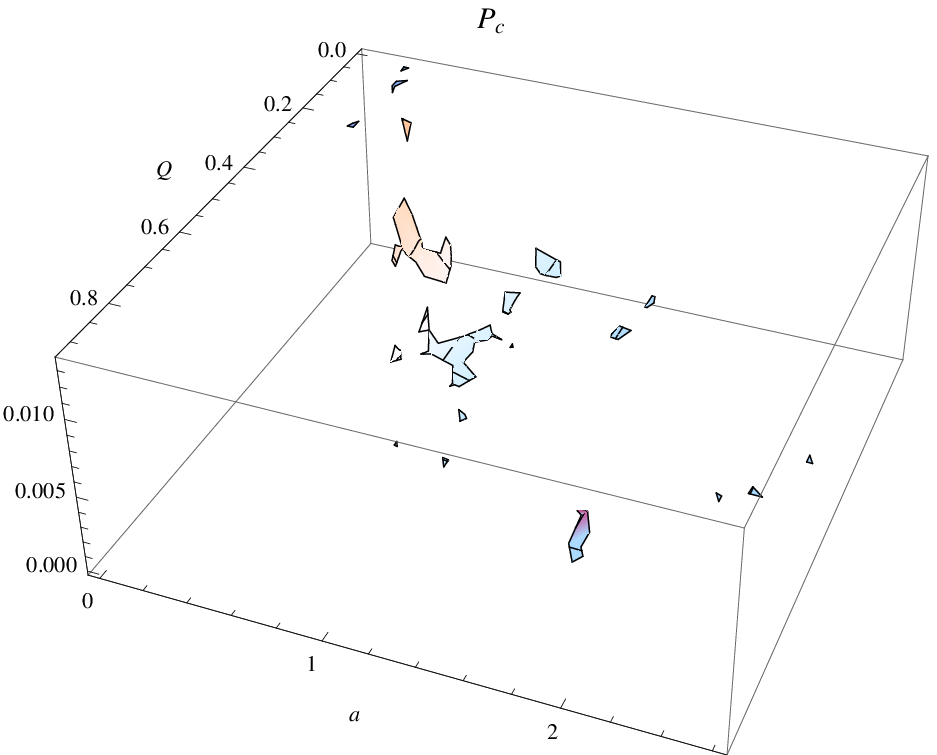}
\end{tabbing}
\caption{The critical temperature and pressure  in the  $(Q, a)$
moduli space } \label{fig1:fig2}

\end{figure}
\end{center}
 From  figure $3$, we observe that the critical pressure is not well
defined in  all regions of the $(Q, a)$ moduli space. For this reason,
we plot  the regions in which the coordinate of the  critical point
is allowed physically. These regions  are  plotted in figure 4.
 \begin{center}
\begin{figure}[!h]
\center
\includegraphics[scale=1]{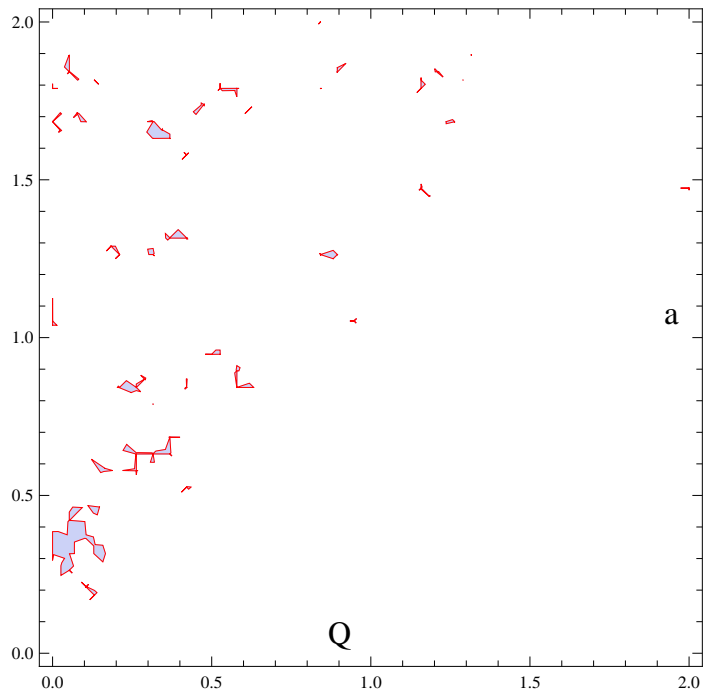}  
\caption{Regions where the coordinates of the critical points are
reals in the  $(Q,a)$ moduli space } \label{fig1:fig2}
\end{figure}
\end{center}

According to the region of the  $(Q, a)$ modul space,  we distinguish
 several  cases. More precisely,      non  black hole transition has
 been observed in the region defined by   $Q>1$  and  $a<1$. However,  in the
region   defined by  $0<Q<1$  and $0<a<2$,  a black hole transition
can be produced for  very small
  $T_c$ and $P_c$.  Moreover,  it has been shown that the black  hole transition can be easily produced for   non
vanishing values of  the charge $Q$ and the parameter $a$.

 In the end
of this work, it is interesting to note that  the extra constraints
on the $(Q,a)$  moduli  space  can be fixed  by calculating the
entropy function.  The  physical conditions, derived from the
thermodynamical principals,   on such a function raise new
constraints on the $(Q, a)$ moduli space. To see that,  we consider
the following action \cite{32}
\begin{equation}
\mathcal{I}=\frac{\beta}{4 G \Xi  \ell ^2}  \left(-r_+^3+\Xi   \ell ^2r_+ +\frac{\ell ^2
   \left(a^2+Q^2\right)}{r_+}+2\frac{  \ell ^2 Q^2 r_+}{a^2+r_+^2}\right)
\end{equation}
It is known that this action  corresponds  to the  Gibbs free energy
\begin{equation}
G=G(P,T)=\frac{1}{4} r_+ \left(1-\frac{6 Q^2}{\left(8 \pi  a^2 P-3\right)
   \left(a^2+r_+^2\right)}\right)-\frac{3 \left(a^2+Q^2\right)}{4 r_+ \left(8
   \pi  a^2 P-3\right)}+\frac{2 \pi  P r_+^3}{8 \pi  a^2 P-3}
\end{equation}
where  $r_+$  can be  understood as a function of the  pressure and
the  temperature, $r_+=r_+(P,T)$, via the state  equation
$(\ref{state})$.  Roughly speaking, the free energy read as,
\begin{eqnarray}
F(T,V)=G-PV=
\frac{r_+^2}{4 r_+} \left(1-\frac{2 Q^2}
{\left(a^2+r_+^2\right) \left(a^2 X-1\right)}\right)+\frac{a^2+Q^2}{1-a^2 X}
+r_+^4 X \left(\frac{1}{a^2 X-1}-2\right)
\end{eqnarray}
where
\begin{equation}
X=\frac{r_+ \left(4 \pi  a^2 T+r_+ \left(4 \pi  r_+ T-1\right)\right)+a^2+Q^2}{r_+^2 \left(a^2+3 r_+^2\right)}
\end{equation}
Based on these quantities, we get  the entropy function
\begin{equation}
S(T,V)=\frac{\pi  \left(r_+^6 \left(2 a^2 X \left(a^2 X-2\right)+3\right)+a^2 r_+^4 \left(2 a^2 X \left(a^2 X-2\right)+3\right)-r_+^2 \left(a^4+3 a^2 Q^2\right)-a^4
   \left(a^2+Q^2\right)\right)}{r_+^2 \left(a^2+3 r_+^2\right) \left(a^2 X-1\right)^2}
\end{equation}
In the  vanishing limit of the $a$  parameter, the entropy function
becomes
\begin{equation}
S(T,V)|_{a=0}=\pi r_+^2.
\end{equation}
Based on the above  formula of the entropy,  we can find the
following constraint, for $\ell=\sqrt{3}$ and $r_+=1$
\begin{equation}
Q\in \mathcal{R},\;\; and \quad a <\frac{1}{\sqrt{14}}\sqrt{-9
Q^2+\sqrt{81 Q^4-36 Q^2+256}+2}.
\end{equation}
These constraints derived from   the entropy  agree  with the
previous ones obtained from the critical values of $T_c$  and  $P_c$.
\\

As a conclusion, in  this work we  have studied the critical behavior
of  the Kerr Newmann AdS black holes in four dimensions  showing
similarities with  the  Van der Waals  gaz in the asymptotic limit.
In particular, we have derived the  coordinates of the critical
points, then we have discussed  the the effect of the $(Q, a)$ space
moduli on such points. As a by product, we have worked out  a universal
number depending on the $(Q, a)$ moduli space and recovered
 the usual value $\frac{3}{8}$ for Reissner Nordstrom-AdS black hole.

\end{document}